\documentclass[graybox]{svmult}

\usepackage{mathptmx}       
\usepackage{helvet}         
\usepackage{courier}        
\usepackage{type1cm}        
\usepackage{makeidx}         
\usepackage{graphicx}        
\usepackage{multicol}        
\usepackage[bottom]{footmisc}

\newcommand\rf[1]{(\ref{eq:#1})}
\newcommand\lab[1]{\label{eq:#1}}

\newcommand\br{\begin{eqnarray}}
\newcommand\er{\end{eqnarray}}
\newcommand\be{\begin{equation}}
\newcommand\ee{\end{equation}}

\newcommand\foot[1]{\footnotemark\footnotetext{#1}}
\newcommand\lb{\lbrack}
\newcommand\rb{\rbrack}

\renewcommand\({\left(}
\renewcommand\){\right)}

\newcommand\bc{\begin{center}}
\newcommand\ec{\end{center}}

\newcommand\partder[2]{\frac{{\partial {#1}}}{{\partial {#2}}}}

\renewcommand\a{\alpha}

\renewcommand\d{\delta}
\newcommand\eps{\epsilon}
\newcommand\vareps{\varepsilon}

\newcommand\G{\Gamma}

\newcommand\h{\frac{1}{2}}
\renewcommand\k{\kappa}
\renewcommand\l{\lambda}
\renewcommand\L{\Lambda}
\newcommand\m{\mu}
\newcommand\n{\nu}

\newcommand\vp{\varphi}
\renewcommand\P{\Phi}
\newcommand\pa{\partial}

\newcommand\pr{\prime}

\renewcommand\th{\theta}

\newcommand\wti{\widetilde}

\newcommand\cB{{\mathcal B}}

\newcommand{\ct}[1]{\cite{#1}}
\newcommand{\bib}[1]{\bibitem{#1}}

\newcommand\PRL[3]{\textsl{Phys. Rev. Lett.} \textbf{#1} (#2) #3}

\newcommand\PRD[3]{\textsl{Phys. Rev.} \textbf{D#1} (#2) #3}

\newcommand\PLB[3]{\textsl{Phys. Lett.} \textbf{#1B} (#2) #3}
\newcommand\CQG[3]{\textsl{Class. Quantum Grav.} \textbf{#1} (#2) #3}

\newcommand\IJMPA[3]{\textsl{Int. J. Mod. Phys.} \textbf{A#1} (#2) #3}

\newcommand\MPLA[3]{\textsl{Mod. Phys. Lett.} \textbf{A#1} (#2) #3}

\newcommand\vpdot{\stackrel{.}{\varphi}}
\newcommand\vpddot{\stackrel{..}{\varphi}}
\newcommand\phidot{\stackrel{.}{\phi}}

\begin{document}

\title*{Metric-Independent Spacetime Volume-Forms and Dark Energy/Dark Matter Unification}
\author{Eduardo Guendelman, Emil Nissimov and Svetlana Pacheva}
\institute{Eduardo Guendelman \at Department of Physics, Ben-Gurion University of
the Negev, Beer-Sheva, Israel \email{guendel@bgu.ac.il}
\and Emil Nissimov \at Institute for Nuclear Research and Nuclear Energy,
Bulgarian Academy of Sciences, Sofia, Bulgaria \email{nissimov@inrne.bas.bg}
\and Svetlana Pacheva \at Institute for Nuclear Research and Nuclear Energy,
Bulgarian Academy of Sciences, Sofia, Bulgaria \email{svetlana@inrne.bas.bg}}
%
%
\maketitle

\abstract{The method of non-Riemannian (metric-independent) spacetime volume-forms 
(alternative generally-covariant integration measure densities) is applied
to construct a modified model of gravity coupled to a single scalar field 
providing an explicit unification of dark energy (as a dynamically generated
cosmological constant) and dust fluid dark matter flowing along geodesics
as an exact sum of two separate terms in the scalar field energy-momentum
tensor. The fundamental reason for the dark species unification is the
presence of a non-Riemannian volume-form in the scalar field action which
both triggers the dynamical generation of the cosmological constant as well
as gives rise to a hidden nonlinear Noether symmetry underlying the dust dark
matter fluid nature. Upon adding appropriate perturbation breaking the
hidden ``dust'' Noether symmetry we preserve the geodesic flow property of
the dark matter while we suggest a way to get growing dark energy in the
present universe' epoch free of evolution pathologies. Also, 
an intrinsic relation between the above modified gravity + single scalar field
model and a special quadratic purely kinetic ``k-essence'' model is
established as a weak-versus-strong-coupling duality.}

\section{Introduction}
\label{intro}

According to the standard cosmological model ($\L$CDM model 
\ct{Lambda-CDM-1}-\ct{Lambda-CDM-3}) the energy density of the late time Universe 
is dominated by two ``dark'' components - around 70 \% made out of ``dark energy''
\ct{dark-energy-observ-1}-\ct{dark-energy-observ-3} and around 25 \% made out of 
``dark matter'' \ct{dark-matter-rev-1}-\ct{dark-matter-rev-3}. Since more
than a decade a principal challenge in modern cosmology is to understand
theoretically from first principles the
nature of both ``dark'' species of the universe's substance as a
manifestation of the dynamics of a single entity of matter. Among the
multitude of approaches to this seminal problem proposed so far are the
(generalized) ``Chaplygin gas'' models \ct{chaplygin-1}-\ct{chaplygin-4}, 
the ``purely kinetic k-essence'' models \ct{scherrer}-\ct{pure-k-essence-3}
based on the class of kinetic ``quintessence'' models  
\ct{k-essence-1}-\ct{k-essence-4}, and more recently -- the so called
``mimetic'' dark matter model \ct{mimetic-grav-1,mimetic-grav-2} and its
extensions \ct{mimetic-grav-extend-1,mimetic-grav-extend-2},
as well as constant-pressure-ansatz models \ct{cruz-etal}.

Here we will describe a new approach achieving unified description of dark
energy and dark matter based on a class of generalized models of gravity
interacting with a single scalar field employing the method of 
non-Riemannian volume-forms on the pertinent spacetime manifold
\ct{TMT-orig-0}-\ct{TMT-orig-4} (for further developments, see 
Refs.\ct{TMT-recent-1,TMT-recent-2}). Non-Riemannian spacetime volume-forms
or, equivalently, alternative generally covariant integration measure densities
are defined in terms of auxiliary maximal-rank antisymmetric tensor gauge fields 
(``measure gauge fields'') unlike the standard Riemannian integration
measure density given given in terms of the square root of the determinant of the 
spacetime metric. These non-Riemannian-measure-modified gravity-matter models
are also called ``two-measure gravity theories''.  

Let us particularly stress that the method of non-Riemannian spacetime
volume-forms is a very powerful one having profound impact in any (field theory) 
models with general coordinate reparametrization invariance, 
such as general relativity and its extensions \ct{TMT-orig-0}-\ct{TMT-orig-4}, 
\ct{TMT-recent-1}-\ct{dusty}, \ct{emergent}-\ct{buggy}; 
strings and (higher-dimensional) membranes \ct{mstring-1,mstring-2}; and supergravity
\ct{susyssb-1,susyssb-2}. Among its main features we should mention:

\begin{itemize}
\item 
Dynamical generation of cosmological constant as arbitrary integration
constant in the solution of the equations of motion for the auxiliary
``measure'' gauge fields (see also Eq.\rf{L-const} below). 
\item
Using the canonical Hamiltonian formalism for Dirac-constrained
systems we find that the auxiliary ``measure'' gauge fields are in fact
almost pure gauge degrees of freedom except for the above mentioned
arbitrary integration constants which are identified with the conserved
Dirac-constrained canonical momenta conjugated to the ``magnetic'' components 
of the ``measure'' gauge fields \ct{quintess,buggy}.
\item
Applying the non-Riemannian volume-form formalism to minimal $N=1$ 
supergravity the appearance of a dynamically generated cosmological constant 
triggers spontaneous supersymmetry breaking and mass generation for the gravitino 
(supersymmetric Brout-Englert-Higgs effect) \ct{susyssb-1,susyssb-2}. Applying the same 
formalism to anti-de Sitter supergravity allows to produce
simultaneously a very large physical gravitino mass and a very small 
{\em positive} observable cosmological constant \ct{susyssb-1,susyssb-2} in accordance 
with modern cosmological scenarios for slowly expanding universe of the present epoch
\ct{dark-energy-observ-1,dark-energy-observ-2,dark-energy-observ-3}. 
\item
Employing two independent non-Riemannian volume-forms produces
effective scalar potential with two infinitely large flat regions
\ct{emergent,quintess}
(one for large negative and another one for large positive values of the scalar 
field $\vp$) with vastly different scales appropriate for a unified description 
of both the early and late universe' evolution.
A remarkable feature is the existence of a stable initial phase of
{\em non-singular} universe creation preceding the inflationary phase
-- stable ``emergent universe'' without ``Big-Bang'' \ct{emergent}.
\end{itemize}

In Section 2 below we briefly discuss a non-standard model of
gravity interacting with a single scalar field which couples symmetrically
to a standard Riemannian as well as to another non-Riemannian volume form
(spacetime integration measure density). We show that the auxiliary
``measure'' gauge field dynamics produces an arbitrary integration constant
identified as a dynamically generated cosmological constant giving rise to a
the dark energy term in the pertinent energy-momentum tensor.
Simultaneously, a hidden strongly nonlinear Noether symmetry of the scalar
Lagrangian action is revealed leading to a ``dust'' fluid representation of
the second term in the energy-momentum tensor, which accordingly is identified
as a ``dust'' dark matter flowing along geodesics. 
Thus, both ``dark'' species are explicitly
unified as an exact sum of two separate contributions to the energy-momentum tensor.

In Section 3 some implications for cosmology are briefly considered.
Specifically, we briefly study an appropriate perturbation of our 
modified-measure gravity + scalar-field model which breaks the above 
crucial hidden Noether symmetry and introduces exchange between the dark 
energy and dark matter components, while preserving the geodesic flow
property of the dark matter fluid. Further, we suggest how to obtain 
a growing dark energy in the present day universe' epoch without invoking any
pathologies of ``cosmic doomsday'' or future singularities kind 
\ct{doomsday-1}-\ct{doomsday-3}.

In Sections 4 below we couple the above modified-measure scalar-field model
to a quadratic $f(R)$-gravity. We derive the pertinent ``Einstein''-frame
effective theory which turns out be a very special quadratic purely kinetic 
``k-essence'' gravity-matter model. The main result here is establishing duality
(in the standard sense of weak versus strong coupling) between the latter and
the original quadratic $f(R)$-gravity plus modified-measure scalar-field model, 
whose matter part delivers an exact unified description of dynamical dark energy 
and dust fluid dark matter.

Section 5 contains our concluding remarks.

For further details, in particular, canonical Hamiltonian treatment and
Wheeler-DeWitt quantization of the above unified model of dark energy and
dark matter, see Refs.\ct{dusty,dusty-2}.


\section{Gravity-Matter Theory With a Non-Riemannian Volume-Form in the
Scalar Field Action -- Hidden Noether Symmetry and Unification of Dark
Energy and Dark Matter}
\label{TMT}

Let us consider the following simple particular case of a non-conventional 
gravity-scalar-field action -- a member of the general class of the 
``two-measure'' gravity-matter theories \ct{TMT-orig-1}-\ct{TMT-orig-4}
(for simplicity we use units with the Newton constant $G_N = 1/16\pi$):
\be
S = \int d^4 x \sqrt{-g}\, R +
\int d^4 x \bigl(\sqrt{-g}+\P(B)\bigr) L(\vp,X) \; .
\lab{TMT-0}
\ee
Here $R$ denotes the standard Riemannian scalar curvature for the pertinent
Riemannian metric $g_{\m\n}$. 
The second term in \rf{TMT-0} -- the scalar field action is constructed in
terms of two mutually independent spacetime volume-forms (integration
measure densities):

(a) $\sqrt{-g} \equiv \sqrt{-\det\Vert g_{\m\n}\Vert}$ is the standard
Riemannian integration measure density;

(b) $\P(B)$ denotes an alternative non-Riemannian generally covariant
integration measure density independent of $g_{\m\n}$ and defining an 
alternative non-Riemannian volume-form:
\be
\P(B) = \frac{1}{3!}\vareps^{\m\n\k\l} \pa_\m B_{\n\k\l} \; ,
\lab{mod-measure}
\ee
where $B_{\m\n\l}$ is an auxiliary maximal rank antisymmetric tensor gauge
field independent of the Riemannian metric, also called ``measure gauge field''.

$L(\vp,X)$ is general-coordinate invariant Lagrangian of a single scalar field 
$\vp (x)$, the simplest example being:
\be
L(\vp,X) = X - V(\vp) \quad ,\quad X \equiv - \h g^{\m\n}\pa_\m \vp \pa_\n \vp \; ,
\lab{standard-L}
\ee
As it will become clear below, the final result about the unification of
dark energy and dark matter resulting from an underlying hidden Noether
symmetry (see \rf{hidden-sym} below) of the scalar field action 
(second term in \rf{TMT-0}) does {\em not} depend on the detailed form of 
$L(\vp,X)$ which could be of an arbitrary 
generic ``k-essence'' form \ct{k-essence-1}-\ct{k-essence-4}:
\be
L(\vp,X) = \sum_{n=1}^N A_n (\vp) X^n - V(\vp) \; ,
\lab{k-essence-L}
\ee
\textsl{i.e.}, a nonlinear (in general) function of the scalar kinetic term $X$.

Due to general-coordinate invariance we have covariant conservation of the 
scalar field energy-momentum tensor:
\be
T_{\m\n} = g_{\m\n} L(\vp,X) + 
\Bigl( 1+\frac{\P(B)}{\sqrt{-g}}\Bigr) \partder{L}{X} \pa_\m \vp\, \pa_\n \vp 
\quad ,\quad \nabla^\n T_{\m\n} = 0 \quad\quad \; .
\lab{EM-tensor}
\ee

Equivalently, energy-momentum conservation \rf{EM-tensor} follows from the
second-order equation of motion w.r.t. $\vp$. The latter, however, becomes
redundant because the modified-measure scalar field action 
(second term in \rf{TMT-0}) exhibits a crucial new property -- it yields a 
{\em dynamical constraint} on $L(\vp,X)$ as a result of the equations of 
motion w.r.t. ``measure'' gauge field $B_{\m\n\l}$:
\be
\pa_\m L (\vp,X) = 0 \;\;\; \longrightarrow \;\;\;
L (\vp,X) = - 2M = {\rm const} \; ,
\lab{L-const}
\ee
in particular, for \rf{standard-L}:
\be
X - V(\vp) = - 2M \;\; \longrightarrow \;\; X = V(\vp) - 2M \; ,
\lab{X-constr}
\ee
where $M$ is arbitrary integration constant. The factor $2$ in front of $M$ is 
for later convenience, moreover, we will take $M>0$ in view of its 
interpretation as a dynamically generated cosmological constant\foot{The physical 
meaning of the ``measure'' gauge field $B_{\m\n\l}$ \rf{mod-measure}
as well as the meaning of the integration constant  $M$ are most straightforwardly 
seen within the canonical Hamiltonian treatment of \rf{TMT-0} 
\ct{dusty}. For more details about the canonical Hamiltonian treatment of general 
gravity-matter theories with (several independent) non-Riemannian volume-forms 
we refer to \ct{quintess,buggy}.}. Indeed, taking into account \rf{L-const},
the expression \rf{EM-tensor} becomes: 
\be
T_{\m\n} = - 2M g_{\m\n} + 
\Bigl( 1+\frac{\P(B)}{\sqrt{-g}}\Bigr)\partder{L}{X} \pa_\m \vp \pa_\n \vp \; .
\lab{EM-tensor-0}
\ee

As already shown in Ref.\ct{dusty} the scalar field action in \rf{TMT-0}
possesses a hidden strongly nonlinear Noether symmetry, namely
\rf{TMT-0} is invariant (up to a total derivative) under the
following nonlinear symmetry transformations:
\be
\d_\eps \vp = \eps \sqrt{X} \quad ,\quad \d_\eps g_{\m\n} = 0 \quad ,\quad
\d_\eps \cB^\m = - \eps \frac{1}{2\sqrt{X}} g^{\m\n}\pa_\n \vp 
\bigl(\P(B) + \sqrt{-g}\bigr)  \; ,
\lab{hidden-sym}
\ee
where $\cB^\m \equiv \frac{1}{3!} \vareps^{\m\n\k\l} B_{\n\k\l}$.
Under \rf{hidden-sym} the action \rf{TMT-0} transforms as \\
$\d_\eps S = \int d^4 x \pa_\m \bigl( L(\vp,X) \d_\eps \cB^\m \bigr)$. 
Then, the standard Noether procedure yields the conserved current:
\be
\nabla_\m J^\m = 0 \quad ,\quad
J^\m \equiv \Bigl(1+\frac{\P(B)}{\sqrt{-g}}\Bigr)\sqrt{2X} 
g^{\m\n}\pa_\n \vp \partder{L}{X} \; .
\lab{J-conserv}
\ee

$T_{\m\n}$ \rf{EM-tensor-0} and $J^\m$ \rf{J-conserv} can be cast into
a relativistic hydrodynamical form: 
\be
T_{\m\n} = - 2M g_{\m\n} + \rho_0 u_\m u_\n \quad ,\quad J^\m = \rho_0 u^\m \; ,
\lab{T-J-hydro}
\ee
where:
\be
\rho_0 \equiv \Bigl(1+\frac{\P(B)}{\sqrt{-g}}\Bigr)\, 2X \partder{L}{X} 
\;\;\; ,\;\; 
u_\m \equiv \frac{\pa_\m \vp}{\sqrt{2X}} \;\; ,\;\; 
 u^\m u_\m = -1 \; .
\lab{rho-0-def}
\ee
For the pressure $p$ and energy density $\rho$ we have accordingly
(with $\rho_0$ as in \rf{rho-0-def}):
\be
p = - 2M = {\rm const} \quad ,\quad
\rho = \rho_0 - p = \Bigl(1+\frac{\P(B)}{\sqrt{-g}}\Bigr)\, 2X \partder{L}{X}
+ 2M \; ,
\lab{p-rho-def}
\ee
where the integration constant $M$ 
appears as {\em dynamically generated cosmological constant}.

Thus, $T_{\m\n}$ \rf{T-J-hydro} represents an exact sum of two contributions of
the two dark species with 
$p = p_{\rm DE} + p_{\rm DM}$ and $\rho = \rho_{\rm DE} + \rho_{\rm DM}$:
\be
p_{\rm DE} = -2M\;\; ,\;\; \rho_{\rm DE} = 2M \quad ; \quad
p_{\rm DM} = 0\;\; ,\;\; \rho_{\rm DM} = \rho_0 \; ,
\lab{DE+DM}
\ee
\textsl{i.e.}, the dark matter component is a dust fluid ($p_{\rm DM} = 0$).

Covariant conservation of $T_{\m\n}$ \rf{T-J-hydro} immediately implies 
{\em both} (i) the covariant conservation of $J^\m = \rho_0 u^\m$ \rf{J-conserv} 
describing dust dark matter ``particle number'' conservation, and (ii)
the geodesic flow equation of the dust dark matter fluid:
\be
\nabla_\m \bigl(\rho_0 u^\m\bigr) = 0 \quad ,\quad 
u_\n \nabla^\n u_\m = 0 \; .
\lab{dust-geo}
\ee


\section{Some Cosmological Implications}
\label{cosmolog}

Let us now consider a perturbation of the initial modified-measure
gravity + scalar-field action \rf{TMT-0} by some additional scalar field
Lagrangian ${\widehat L}(\vp,X)$ independent of the initial scalar Lagrangian 
$L(\vp,X)$:
\be
{\widehat S} = \int d^4 x \sqrt{-g}\, R 
+ \int d^4 x \bigl(\sqrt{-g}+\P(B)\bigr) L(\vp,X) 
+ \int d^4 x \sqrt{-g}\, {\widehat L}(\vp,X) \; .
\lab{TMT-perturb}
\ee
An important property of the perturbed action \rf{TMT-perturb} is that
once again the scalar field $\vp$-dynamics is given by the unperturbed dynamical 
constraint Eq.\rf{L-const} of the initial scalar Lagrangian $L(\vp,X)$, which is
completely independent of the perturbing scalar Lagrangian ${\widehat L}(\vp,X)$.

Henceforth, for simplicity we will take the scalar Lagrangians in the
canonical form $L(\vp,X) = X - V(\vp) \; ,\; {\widehat L}(\vp,X) = X - U(\vp)$,
where $U(\vp)$ is independent of $V(\vp)$.

The associated scalar field energy-momentum tensor now reads (cf.
Eqs.\rf{T-J-hydro}-\rf{p-rho-def}):
\be
{\widehat T}_{\m\n} =  {\widehat \rho_0} u_\m u_\n + 
g_{\m\n}\bigl(-4M+V-U\bigr)  \quad ,\quad
{\widehat \rho_0} \equiv 2(V-2M)\Bigl( 1 + \frac{\P(B)}{\sqrt{-g}}\Bigr) \; ,
\lab{EM-tensor-perturb}
\ee
or, equivalently:
\br
{\widehat T}_{\m\n} = \bigl({\widehat \rho} + {\widehat p} \bigr) u_\m u_\n + 
{\widehat p}\, g_{\m\n} \quad ,\quad
{\widehat p} = -4M + V - U \; ,
\lab{EM-tensor-perturb-0} \\
{\widehat \rho} = {\widehat \rho}_0 - {\widehat p}
= 2(V-2M)\Bigl( 1 + \frac{\P(B)}{\sqrt{-g}}\Bigr)  + 4M + U - V \; ,
\lab{rho-def-perturb}
\er
where \rf{X-constr} is used. 

The perturbed energy-momentum \rf{EM-tensor-perturb} conservation 
$\nabla^\m {\widehat T}_{\m\n} = 0$ now implies:

\begin{itemize}
\item
The perturbed action \rf{TMT-perturb} does not any 
more possess the hidden symmetry \rf{hidden-sym} and, therefore, the conservation
of the dust particle number current $J^\m = \rho_0 u^\m$ \rf{T-J-hydro} is now 
replaced by:
\be
\nabla^\m \bigl({\widehat \rho}_0 u_\m\bigr)
+ \sqrt{2(V-2M)}\Bigl(\partder{V}{\vp} - \partder{U}{\vp}\Bigr) = 0 \; .
\lab{hydro-perturb}
\ee
\item
{\em Once again} we obtain the geodesic flow equation for the dark matter ``fluid''
(second Eq. \rf{dust-geo}).
Let us stress that this is due to the fact that the perturbed pressure 
${\widehat p}$ (second relation in \rf{EM-tensor-perturb-0}), because of the 
dynamical constraint \rf{X-constr} triggered by the non-Riemannian volume-form 
in \rf{TMT-perturb}, is a function of $\vp$ only but not of $X$.
\end{itemize}

Thus, we conclude that the geodesic flow dynamics of the cosmological fluid
described by the action \rf{TMT-perturb} persist irrespective of the
presence of the perturbation (last term in \rf{TMT-perturb}) as well as of the 
specific form of the latter.

In the cosmological context, when taking the spacetime metric in the standard
Friedmann-Lemaitre-Robertson-Walker (FLRW) form, the scalar field is assumed
to be time-dependent only: $\vp = \vp(t)$. Thus, in this case the dynamical
constraint Eq.\rf{X-constr} and its solution assume the form:
\be
\vpdot^2 = 2\bigl( V(\vp) - 2M\bigr) \;\; \longrightarrow \;\;
\int_{\vp (0)}^{\vp (t)} \frac{d\vp}{\sqrt{2\bigl( V(\vp) - 2M\bigr)}} = \pm t \; . 
\lab{vp-eq-2-0} 
\ee
Choosing the $+$ sign in \rf{vp-eq-2-0} corresponds to $\vp (t)$ monotonically 
growing with $t$ irrespective of the detailed form of the potential $V(\vp)$.
The only condition due to consistency of the dynamical constraint 
(first Eq.\rf{vp-eq-2-0}) is $V(\vp) > 2M$ for the whole interval of
classically accessible values of $\vp$. Also, note the ``strange'' looking
second-order (in time derivatives) form of the first Eq.\rf{vp-eq-2-0}:
$\vpddot - \pa V/\pa\vp = 0$ ,
where we specifically stress on the {\em opposite} sign in the force term.
Thus, it is fully consistent for $\vp (t)$ to ``climb'' a growing w.r.t. $\vp$ 
scalar potential.

As already stressed above, the dynamics of the $\vp (t)$ does not depend
at all on the presence of the perturbing scalar potential $U(\vp)$.
Therefore, if we choose the perturbation $U(\vp)$ in \rf{TMT-perturb} such that
the potential difference $U(\vp)-V(\vp)$ is a growing function at large $\vp$ 
(\textsl{e.g.}, $U(\vp)-V(\vp) \sim e^{\a\vp}$, $\a$ small positive) then, when $\vp (t)$
evolves through \rf{vp-eq-2-0} to large positive values, it (slowly) ``climbs'' 
$U(\vp)-V(\vp)$ and according to the expression 
${\widehat \rho}_{DE} = 4M + U(\vp)-V(\vp)= - {\widehat p}$ 
for the dark energy density (cf. \rf{EM-tensor-perturb}-\rf{EM-tensor-perturb-0}), 
the latter will (slowly) grow up! Let us emphasize that in this way we obtain growing
dark energy of the ``late'' universe without any pathologies in the 
universe' evolution like ``cosmic doomsday'' or future singularities 
\ct{doomsday-1}-\ct{doomsday-3}.


\section{Duality to Purely Kinetic ``K-Essence''}
\label{duality}

Let us now consider a different perturbation of the modified-measure
gravity + scalar-field action \rf{TMT-0} by replacing the standard
Einstein-Hilbert gravity action (the first term in \rf{TMT-0}) with a
$f(R)=R-\a R^2$ extended gravity action in the first-order Palatini formalism:
\be
S^{(\a)} = \int d^4 x \sqrt{-g} \bigl( R(g,\G) - \a R^2(g,\G)\bigr)
+ \int d^4 x \bigl(\sqrt{-g}+\P(B)\bigr) L(\vp,X) \; ,
\lab{TMT-A}
\ee
where $R(g,\G) = g^{\m\n} R_{\m\n}(\G)$,
\textsl{i.e.}, with {\em a priori} independent metric $g_{\m\n}$ and affine 
connection $\G^\m_{\n\l}$.

Since the scalar field action -- the second term in \rf{TMT-A} --
remains the same as in the original action \rf{TMT-0}, and the hidden
nonlinear Noether symmetry \rf{hidden-sym} does not affect the metric, all results in
Section 2 remain valid. Namely, the Noether symmetry \rf{hidden-sym} 
produces ``dust'' fluid particle number conserved current 
(first Eq.\rf{dust-geo})
and interpretation of $\vp$ as describing simultaneously dark energy 
(because of the dynamical scalar Lagrangian constraint \rf{L-const})
and dust dark matter with geodesic dust fluid flow (second Eq.\rf{dust-geo}) remains intact.

However, the gravitational equations of motion derived from \rf{TMT-A} are
not of the standard Einstein form:
\be
R_{\m\n}(\G) = \frac{1}{2 f^{\pr}_R} \bigl\lb T_{\m\n} + f(R) g_{\m\n} \bigr\rb \; ,
\lab{non-einstein-eqs} 
\ee
where $f(R) = R(g,\G ) - \a R^2(g,\G ) \; ,\; f^{\pr}_R = 1 - 2\a R(g,\G )$
and $T_{\m\n}$ is the same as in \rf{EM-tensor-0}.

The equations of motion w.r.t. independent $\G^\m_{\n\l}$ resulting from 
\rf{TMT-A} yield (for an analogous derivation, see \ct{TMT-orig-1}) the following
solution for $\G^\m_{\n\l}$ as a Levi-Civita connection:
\be
\G^\m_{\n\l} = \G^\m_{\n\l}({\overline g}) = 
\h {\overline g}^{\m\k}\(\pa_\n {\overline g}_{\l\k} + \pa_\l {\overline g}_{\n\k} 
- \pa_\k {\overline g}_{\n\l}\) \; ,
\lab{G-eq}
\ee
w.r.t. to the Weyl-rescaled metric ${\overline g}_{\m\n}$:
\be
{\overline g}_{\m\n} = f^{\pr}_R\, g_{\m\n} \; ,
\lab{bar-g}
\ee
so that ${\overline g}_{\m\n}$ is called (physical) ``Einstein-frame'' metric.
In passing over to the ``Einstein-frame'' it is also useful to perform
the following $\vp$-field redefinition:
\be
\vp \to {\wti \vp} = \int \frac{d\vp}{\sqrt{\bigl( V(\vp)-2M\bigr)}} 
\quad ,\quad
X \to {\wti X} = -\h {\overline g}^{\m\n} \pa_\m {\wti \vp} \pa_\n {\wti \vp}
= \frac{1}{f^{\pr}_R} \; ,
\lab{new-vp}
\ee
where the last relation follows from the Lagrangian dynamical constraint \rf{X-constr} 
together with \rf{bar-g}.

Derivation of the explicit expressions for the Einstein-frame gravitational
equations, \textsl{i.e.}, equations w.r.t. Einstein-frame metric \rf{bar-g}
and the Einstein-frame scalar field (first Eq.\rf{new-vp}), 
yields the latter in the standard form of Einstein gravity equations:
\be
{\overline R}_{\m\n} - \h {\overline g}_{\m\n} {\overline R} = 
\h {\overline T}_{\m\n} \; .
\lab{standard-einstein-eqs}
\ee
Here the following notations are used:

(i) ${\overline R}_{\m\n}$ and ${\overline R}$ are the standard Ricci
tensor and scalar curvature of the Einstein-frame metric \rf{bar-g}.

(ii) The Einstein-frame energy-momentum tensor:
\be
{\overline T}_{\m\n} = {\overline g}_{\m\n} {\overline L}_{\rm eff}
- 2 \partder{{\overline L}_{\rm eff}}{{\overline g}^{\m\n}}
\lab{EM-tensor-eff}
\ee
is given in terms of the following effective ${\wti \vp}$-scalar field Lagrangian 
of a specific quadratic purely kinetic ``k-essence'' form:
\be
{\overline L}_{\rm eff} ({\wti X}) = 
\Bigl(\frac{1}{4\a} - 2M\Bigr) {\wti X}^2 - \frac{1}{2\a} {\wti X} 
+ \frac{1}{4\a} \; .
\lab{L-eff} 
\ee

Thus, the Einstein-frame gravity+scalar-field action reads:
\be
S_{\rm k-ess} = \int d^4 \sqrt{-{\overline g}} \Bigl\lb {\overline R} +
\Bigl(\frac{1}{4\a} - 2M\Bigr) {\wti X}^2 - \frac{1}{2\a} {\wti X} + 
\frac{1}{4\a}\Bigr\rb \; .
\lab{k-essence-A}
\ee

The Einstein-frame effective energy-momentum-tensor \rf{EM-tensor-eff} 
in the perfect fluid representation reads (taking into account 
the explicit form of ${\overline L}_{\rm eff}$ \rf{L-eff}):
\br
{\bar T}_{\m\n} = {\overline g}_{\m\n} {\wti p} + 
{\wti u}_\m {\wti u}_\n \bigl({\wti \rho} + {\wti p}\bigr) \quad ,\quad
{\wti u}_\m \equiv \frac{\pa_\m {\wti \vp}}{\sqrt{2{\wti X}}} \;\; , \;\;
{\overline g}^{\m\n} {\wti u}_\m {\wti u}_\n = - 1 \; ,
\lab{EM-tensor-wti} \\
{\wti p} = \Bigl(\frac{1}{4\a} - 2M\Bigr) {\wti X}^2  
- \frac{1}{2\a} {\wti X} + \frac{1}{4\a} \;\; ,\;\;
{\wti \rho} = 3 \Bigl(\frac{1}{4\a} - 2M\Bigr){\wti X}^2 
- \frac{1}{2\a} {\wti X} - \frac{1}{4\a} \; . 
\lab{wti-p-rho}
\er

Let us stress that the quadratic purely kinetic ``k-essence'' scalar
Lagrangian \rf{L-eff} is indeed a very special one:

\begin{itemize}
\item
The three coupling constants in \rf{L-eff} depend only
on two independent parameters $(\a, M)$, the second one being a dynamically generated
integration constant in the original theory \rf{TMT-A}.
\item
The quadratic gravity term $-\a R^2$ in \rf{TMT-A} is just a small
perturbation w.r.t. the initial action \rf{TMT-0} when $\a \to 0$, whereas 
the coupling constants in the Einstein-frame effective action 
\rf{k-essence-A} diverge as $1/\a$, \textsl{i.e.}, weak coupling in
\rf{TMT-A} is equivalent to a strong coupling in \rf{k-essence-A}.
\item
Due to the apparent Noether symmetry of \rf{L-eff} under
constant shift of ${\wti \vp}$ (${\wti \vp} \to {\wti \vp} + {\rm const}$)
the corresponding Noether conservation law is identical to the 
${\wti \vp}$-equations of motion:
\be
{\overline \nabla}_\m \Bigl({\overline g}^{\m\n} \pa_\n {\wti \vp}
\partder{{\wti L}_{\rm eff}}{{\wti X}}\Bigr) = 0 \; ,
\lab{vp-eqs-wti}
\ee
where ${\overline \nabla}_\m$ is covariant derivative w.r.t. the
Levi-Civita connection \rf{G-eq} in the ${\overline g}_{\m\n}$-(Einstein) frame.
Eq.\rf{vp-eqs-wti} is the Einstein-frame counterpart of the ``dust'' Noether
conservation law \rf{J-conserv} in the original theory \rf{TMT-0} or \rf{TMT-A}.
\end{itemize}

Thus, we have found an explicit duality in the usual sense of ``weak versus
strong coupling'' between the original non-standard 
gravity+scalar-field model providing exact unified description
of dynamical dark energy and dust fluid dark matter in the matter sector, 
on one hand, and a special quadratic purely kinetic ``k-essence'' gravity-matter 
model, on the other hand. The latter dual theory arises as the ``Einstein-frame''
effective theory of its original counterpart.

To make explicit the existence of smooth strong coupling limit $\a \to 0$ 
{\em on-shell} in the dual ``k-essence'' energy density ${\wti\rho}$ and ``k-essence''
pressure ${\wti p}$ \rf{wti-p-rho} in spite of the divergence of the
corresponding constant coefficients, let us consider a reduction of the dual 
quadratic purely kinetic ``k-essence'' 
gravity + scalar-field model \rf{k-essence-A} for the
Friedmann-Lemaitre-Robertson-Walker (FLRW) class of metrics:
\be
ds^2 = - N^2(t) dt^2 + a^2(t) \Bigl\lb \frac{dr^2}{1-K r^2}
+ r^2 (d\th^2 + \sin^2\!\th d\phi^2)\Bigr\rb \; .
\lab{FLRW}
\ee

The FLRW reduction of the $\phi \equiv {\wti \vp}$-equation of motion
\rf{vp-eqs-wti} (using henceforth the gauge $N=1$) reads:
\be
\frac{d p_\phi}{dt} = 0 \;\; \longrightarrow \;\;
p_\phi = a^3 \Bigl\lb - \frac{1}{2\a} \phidot + 
\bigl(\frac{1}{4\a} - 2M\bigr) \phidot^3 \Bigr\rb \; ,
\lab{phi-eq}
\ee
where $p_\phi$ is the constant conserved canonically conjugated momentum 
of $\phi\equiv {\wti \vp}$. Thus, the velocity 
$\phidot = \phidot\!\!(p_\phi/a^3)$ is a function of the Friedmann scale factor
$a(t)$ through the ratio $p_\phi/a^3$ and solves the cubic algebraic equation
\rf{phi-eq} for any $\a$. For small $\a$ we get:
\be
\phidot (p_\phi/a^3) \simeq \sqrt{2} 
+ \a \Bigl(4\sqrt{2}M + \frac{p_\phi}{a^3}\Bigr) + \mathrm{O}(\a^2) \; .
\lab{phidot-small-a}
\ee
Then, inserting \rf{phidot-small-a} into the FLRW-reduced 
${\wti X}=\h \phidot^2$ and substituting it into the expressions
\rf{wti-p-rho} we obtain for the small-$\a$ asymptotics of the ``k-essence'' energy
density and ``k-essence'' pressure:
\br
{\wti \rho} = 2M + \sqrt{2}\frac{p_\phi}{a^3}  
+ \a \Bigl\lb 16M^2 + 4\sqrt{2}M \frac{p_\phi}{a^3} + 
\h \bigl(\frac{p_\phi}{a^3}\bigr)^2 \Bigr\rb + {\rm O}(\a^2) \; ,
\lab{rho-small-alpha} \\
{\wti p} = - 2M - \a \Bigl\lb 16M^2 - 
\h \bigl(\frac{p_\phi}{a^3}\bigr)^2 \Bigr\rb  + {\rm O}(\a^2) \; .
\lab{p-small-alpha}
\er
The limiting values ${\wti \rho} = 2M + \sqrt{2}\frac{p_\phi}{a^3}$ and
${\wti p} = - 2M$ precisely coincide with the corresponding values of
$\rho$ and $p$ \rf{p-rho-def} in the FLRW reduced original theory \rf{TMT-0}
\ct{dusty}.

\section{Conclusions}
\label{conclude}

In the present note we have demonstrated the power of the method of non-Riemannian 
spacetime volume-forms (alternative generally-covariant integration measure 
densities) by applying it to construct a modified model of gravity coupled
to a single scalar field which delivers a unification of dark energy (as a
dynamically generated cosmological constant) and dust fluid dark matter
flowing along geodesics (due to a hidden nonlinear Noether symmetry). 
Both ``dark'' species appear as an
exact sum of two separate contributions in the energy-momentum tensor of the
single scalar field. Upon perturbation of the scalar field action, which
breaks the hidden ``dust'' Noether symmetry but preserves the geodesic flow
property, we show how to obtain a growing dark energy in the late Universe
without evolution pathologies. Furthermore, we have established a duality 
(in the standard
sense of weak versus strong coupling) of the above model unifying dark
energy and dark matter, on one hand, and a specific quadratic purely kinetic
``k-essence'' model. This duality elucidates the ability of purely kinetic
``k-essence'' theories to describe approximately the unification of dark
energy and dark matter and explains how the ``k-essence'' description becomes exact in
the strong coupling limit on the ``k-essence'' side.


\begin{acknowledgement}
We gratefully acknowledge support of our collaboration through the 
academic exchange agreement between the Ben-Gurion University in Beer-Sheva,
Israel, and the Bulgarian Academy of Sciences. 
S.P. and E.N. have received partial support from European COST actions
MP-1210 and MP-1405, respectively, as well from Bulgarian National Science
Fund Grant DFNI-T02/6. 
\end{acknowledgement}
\end{document}